\documentclass[twocolumn]{aastex631}

\usepackage{natbib}
\usepackage{amsmath}
\usepackage{subfigure}
\setcitestyle{sort,comma}
\usepackage{csvsimple}

\usepackage[version=4]{mhchem} 
\usepackage{seqsplit}

\providecommand{\apjl}{ApJL}
\providecommand{\apjs}{ApJS}
\providecommand{\aap}{A\&Aj}

\graphicspath{{./}{figures/}}

\begin{document}


\title{Biases in Exoplanet Transmission Spectra Introduced by Limb Darkening Parametrization}

\newcommand{\umontreal}{Department of Physics and Trottier Institute for Research on Exoplanets, Universit\'{e} de Montr\'{e}al, Montr\'{e}al, QC, Canada \href{mailto:louis-philippe.coulombe@umontreal.ca}{louis-philippe.coulombe@umontreal.ca}}

\author[0000-0002-2195-735X]{Louis-Philippe Coulombe} 
\affil{\umontreal}

\author[0000-0001-6809-3520]{Pierre-Alexis Roy} 
\affil{\umontreal}

\author[0000-0001-5578-1498]{Bj\"{o}rn Benneke} 
\affil{\umontreal}


\begin{abstract}

One of the main endeavors of the field of exoplanetary sciences is the characterization of exoplanet atmospheres on a population level.
The current method of choice to accomplish this task is transmission spectroscopy, where the apparent radius of a transiting exoplanet is measured at multiple wavelengths in search of atomic and molecular absorption features produced by the upper atmosphere constituents.
To extract the planetary radius from a transit light curve, it is necessary to account for the decrease in luminosity away from the center of the projected stellar disk, known as the limb darkening.
Physically-motivated parametrizations of the limb darkening, in particular of the quadratic form, are commonly used in exoplanet transit light-curve fitting. Here, we show that such parametrizations can introduce significant wavelength-dependent biases 
in the transmission spectra currently obtained with all instrument modes of the JWST, and thus have the potential to affect atmospheric inferences.
To avoid such biases, we recommend the use of standard limb-darkening parametrizations with wide uninformative priors that allow for non-physical stellar intensity profiles in the transit fits, and thus for a complete and symmetrical exploration of the parameter space. We further find that fitting the light curves at the native resolution results in errors on the measured transit depths that are significantly smaller compared to light curves that are binned in wavelength before fitting, thus potentially maximizing the amount of information that can be extracted from the data.
\end{abstract}


\section{Introduction}\label{sec:intro}

The transit method is arguably the most powerful tool currently at our disposition to detect exoplanets \citep{Charbonneau_2000} and characterize their atmospheres \citep{Charbonneau_2002}. With the wide wavelength coverage and unprecedented photometric precision of the JWST, our understanding of exoplanet atmospheres through transmission spectroscopy is now rapidly and constantly evolving \citep[e.g.,][and many more]{JWST_ERS_2023,Gao_2023,bell2023methane,madhusudhan2023carbonbearing,Dyrek_2023,Xue_2024,benneke2024jwst,Welbanks_2024,cadieux2024transmissionspectroscopyhabitablezone}. 

As a planet transits its host star from an observer's point of view, the measured stellar flux decreases by a factor of $D\approx (R_\mathrm{p}/R_\star)^2$, enabling the extraction of the planetary radius $R_\mathrm{p}$ from this signal, provided that the stellar radius $R_\star$ is known.
When performed with spectroscopic instruments, this method yields the wavelength-dependent apparent radius of an exoplanet and is highly sensitive to the presence of atomic or molecular absorbers in its atmosphere \citep{Seager_2000}. 

While the depth of a transit is, to first order, dictated by the planetary radius, the exact shape of the light curve also depends on the limb-darkening profile of the host star as well as the orbital parameters of the system \citep{Seager_2003}. The limb darkening describes the non-uniform luminosity from the stellar disk, which appears brighter along an observer's line-of-sight in the center of the disk than near the limbs, whereas the orbital parameters define the path of the transiting planet (transit chord) across said stellar disk. 
The effect of limb darkening is highly wavelength-dependent  due to the different optical depths and effective temperatures probed when observing the center (head-on) or the limb (slanted geometry) of the disk, which result in large center-to-limb contrast at optical wavelengths that gradually decreases towards the infrared. Because of this, limb-darkening coefficients are generally fitted independently at each wavelength when performing transmission spectroscopy.


Several parametrizations of the limb darkening have been proposed thus far, such as: linear \citep{Milne1921}, quadratic \citep{Kopal1950}, non-linear \citep{Claret2000}, square-root \citep{diaz_cordovez_1992}, and exponential \citep{Claret_2003}, with the quadratic limb-darkening law currently being the most widely used in exoplanetary sciences. This parametrization describes the normalized stellar intensity profile $I_\star(\mu)/I_0$ as a quadratic function of the normalized radial coordinate $\mu$:

\begin{equation}
\frac{I_\star(\mu)}{I_0} = 1 - u_1 (1-\mu) - u_2 (1-\mu)^2,
\end{equation}

\medskip
\noindent 
where $u_1$ and $u_2$ are the quadratic limb-darkening coefficients (LDCs). Due to the flexibility of this parametrization, it is possible for a given set of parameters [$u_1$,$u_2$] to produce a stellar intensity profile where the intensity increases towards the limbs (limb brightening) or reaches negative values, scenarios that are non-physical for stars.
To avoid considering such solutions when fitting a transit light curve, constraints and reparametrizations of the LDCs have been proposed \citep[e.g.][]{Burke_2007,Carter2009,Kipping_2013}, with the physically-motivated parametrization of \citet{Kipping_2013} (from hereon simply referred to as [$q_1$,$q_2$]) now commonly used for exoplanet transit light curves. 

Explicitly, the [$q_1$,$q_2$] parametrization is obtained by imposing two physical conditions to the quadratic limb-darkening problem: the stellar intensity profile must be everywhere-positive, and it must decrease monotonically away from the center of the stellar disk \citep{Kipping_2013}. This parametrization is especially useful as it significantly reduces the LDCs parameter space and allows for the uniform sampling of $q_1$ and $q_2$ values over a simple unit square. 

Even though the physical assumptions used to define these reparametrizations of the LDCs are valid, we find that a subtle but significant bias can arise from them: light-curve fits are biased towards smaller planetary radii when the limb-darkening effect becomes negligible. This is most relevant for transmission spectroscopy at the infrared wavelengths probed by the JWST \citep{gardner_james_2006}, the decommissioned Spitzer \citep{werner_spitzer_2004}, and future missions such as the Ariel \citep{pascale_ariel_2018} and Twinkle \citep{Stotesbury_2022} space telescopes. While we specifically focus on the bias introduced by the [$q_1$,$q_2$] parametrization in this work, we note that any prior or parametrization enforcing physical solutions for the stellar intensity profile are prone to such biases. 


In this work, we present, describe, and quantify the bias that arises from physics-driven reparametrizations of the limb-darkening problem when used in exoplanet transit light-curve fitting. In section \ref{sec:methods}, we describe the methods used to quantify the bias on simulated and real observations. We present the results in section \ref{sec:results}, and we discuss the implications for atmospheric inference, as well as present a bias-free reparametrization in section \ref{sec:discu}. Finally, our conclusions are presented in section \ref{sec:conclusions}.

\section{Methods}\label{sec:methods}

In this section, we describe the methodology used to quantify the bias introduced by physically-motivated limb-darkening parametrizations on the measured transit depth. We begin by characterizing this effect on simulated observations and proceed to quantify this effect on the JWST NIRISS/SOSS observations of WASP-39\,b afterward. 

\subsection{Light-curve fitting of simulated data}\label{sec:methods_sim}

We begin this work by performing transit light-curve fitting on simulated observations where the ground truth is known. We simulate transit light curves using the \texttt{batman} python package \citep{Kreidberg_2015} for the case of a planet in a circular orbit ($e$ = 0, $\omega$ = 90$^\circ$) with planet-to-star radius ratio $R_p/R_\star$ = 0.1, semi-major axis $a/R_\star$ = 4, impact parameter $b$ = 0, and orbital period $P$ = 1 day. We simulate the transit assuming no limb-darkening ($u_1$ = 0, $u_2$ = 0), which is analogous to transit observations at infrared wavelengths where the bias on the planetary radius is strongest. The simulated transit light curve consists of a thousand equally-spaced integrations spanning times $t\in[-0.25,0.25]$ days, where $t=0$ is the mid-transit time. We consider error envelopes for the simulated observations rather than adding white noise to the model to avoid the introduction of randomness in the retrieved parameter constraints. For the fits, we consider seven photometric scatter values ranging from 1 to 1000\,ppm with equal spacing in log scale. We note that the values of photometric scatter considered here are for an integration time of 0.72\,minutes and thus these values cannot be directly compared to observations that have a different cadence.

We subsequently fit each light curve twice: fitting for $R_p/R_\star$ ($\mathcal{U}$[0,0.5]), $u_1$ ($\mathcal{U}$[-3,3]) and $u_2$ ($\mathcal{U}$[-3,3]) for the standard LDCs case as well as $R_p/R_\star$, $q_1$ ($\mathcal{U}$[0,1]) and $q_2$ ($\mathcal{U}$[0,1]) for the physically-motivated LDCs case. We consider wide priors for $u_1$ and $u_2$ to ensure that a diverse range of stellar intensity profiles may be modelled, including limb-brightening scenarios. The light curves are fitted considering the standard chi-square likelihood and the parameter space is sampled using the \texttt{emcee} python package \citep{Foreman_Mackey_2013}. We use 4 walkers per free parameter and run the chains for 100,000 steps, tossing the first 60\% of the steps as burn-in.

\begin{figure*}[hbt!]
\begin{center}
\includegraphics[width=\textwidth]{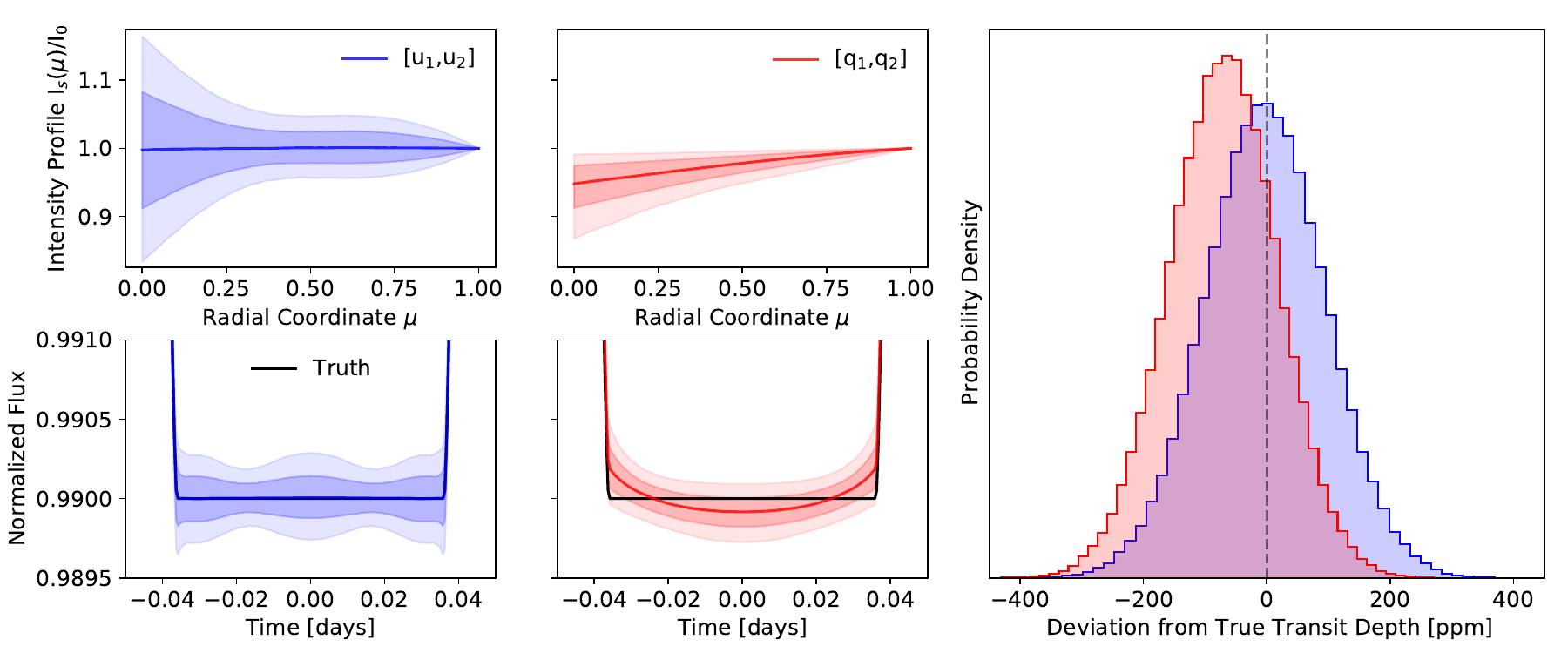}
\end{center}
\vspace{-5mm}\caption{Impact of the choice of limb-darkening on the light curve fit and measured transit depth for the case of no limb darkening and photometric scatter of $\sigma$ = 1000\,ppm. \textbf{Top left:} Median stellar intensity profile $I_\mathrm{s}(\mu)/I_0$ (blue line) from the transit fit to the simulated observations using the [$u_1$,$u_2$] limb-darkening parametrization. The 1-$\sigma$ and 2-$\sigma$ confidence intervals are shown by the shaded regions. \textbf{Bottom left:} Median transit model (blue) obtained from the transit fit with the [$u_1$,$u_2$] limb-darkening parametrization. The 1-$\sigma$ and 2-$\sigma$ confidence intervals are shown by the shaded regions. \textbf{Top middle:} Same as \textbf{Top left} considering the [$q_1$,$q_2$] parametrization. \textbf{Bottom middle:} Same as \textbf{Bottom left} considering the [$q_1$,$q_2$] parametrization. \textbf{Right:} Probability density distributions of the measured transit depth minus the true transit depth for the [$u_1$,$u_2$] (blue) and [$q_1$,$q_2$] (red) parametrizations. The transit depth measured using [$u_1$,$u_2$] is centered on the expected value, whereas the one measured using [$q_1$,$q_2$] peaks at a smaller value.} 
\label{fig:lightcurves}
\end{figure*}

\subsubsection{Light-curve fitting of the NIRISS/SOSS observations of WASP-39\,b}

We proceed to explore the impact of the choice of limb-darkening parametrization on the NIRISS/SOSS spectrum of WASP-39\,b. A transit of WASP-39\,b was observed on 26 July 2022 using the NIRISS instrument \citep{doyon2023near} of the JWST in the SOSS mode \citep{Albert_2023} as part of the Early Release Science (ERS) program \citep{Bean_2018}, the full analysis of which is presented in \citet{Feinstein_2023}. WASP-39\,b is currently the only exoplanet that has been observed in transmission with all instrument modes of the JWST \citep{Ahrer_2023,Alderson_2023,Feinstein_2023,Rustamkulov_2023,powell_sulfur_2024,carter_benchmark_2024}, making it an ideal test case for the exploration of the bias introduced by limb-darkening assumptions. We further choose the NIRISS/SOSS instrument since it covers a large wavelength range (0.85--2.85\,$\mu$m) at a high resolving power (R = 500 -- 1400), and therefore should display a transition from light curves with important limb darkening to light curves where the limb darkening becomes negligible and the bias on the planetary radius arises.

We begin from the raw spectroscopic light curves extracted from the order 1 spectral trace ($\lambda$ = 0.85--2.85\,$\mu$m) using the \texttt{NAMELESS} reduction \citep{Coulombe2023,Radica_2023,Lim2023,Fournier_Tondreau_2023} presented in \citet{Feinstein_2023}. We consider a linear trend for the systematics model with a normalization factor $c$ ($\mathcal{U}$[-10$^9$,10$^9$]) and slope $v$ ($\mathcal{U}$[-10$^9$,10$^9$]). The planetary radius $R_p/R_\star$ and LDCs are kept free for both limb-darkening parametrizations, considering the same priors as in section \ref{sec:methods_sim}. We fix the orbital parameters to $T_0$ = 2459787.556740 BJD, $a/R_\star$ = 11.4, and $b$ = 0.4510 following the values measured from the \texttt{NAMELESS} order 1 white-light curve in \citet{Feinstein_2023}. For both the [$u_1$,$u_2$] and [$q_1$,$q_2$] parametrizations, we fit the light curves at native resolution (i.e., one light curve per detector column) and at a fixed resolving power of R = 80. Each light curve fit is run for 10,000 steps using 4 walkers per free parameter, where we again discard the first 60\% of the steps as burn-in.


\section{Results}\label{sec:results}

\begin{figure}
\begin{center}
\includegraphics[width=0.45\textwidth]{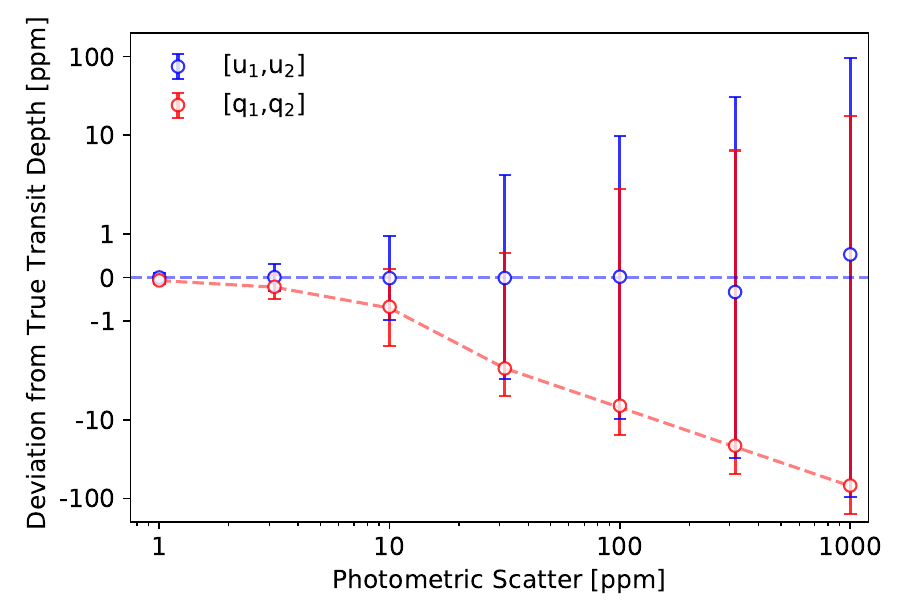}
\end{center}
\vspace{-5mm}\caption{Magnitude of the bias on the measured transit depth as a function of photometric scatter. The data points show the deviation from the true transit depth, along with the 1-$\sigma$ uncertainty, as a function of photometric scatter for both the [$u_1$,$u_2$] and [$q_1$,$q_2$] parametrizations. We use a symmetric log scale for the y-axis. Contrastingly from the [$u_1$,$u_2$] parametrization, the [$q_1$,$q_2$] parametrization shows an increasing bias of the transit depth when going toward larger scatter. The blue dashed line corresponds to no bias on the transit depth, with the measured transit depths obtained using the [$u_1,u_2$] parametrization following that trend. The bias on the transit depth when using the [$q_1$,$q_2$] parametrization closely follows the red dashed line, which traces the trend $y = - \sqrt{2/\pi}~\sigma_{t,d}$, where $\sigma_{t,d}$ is the measured transit depth uncertainty.} 
\label{fig:scatter}
\end{figure}

\begin{figure*}
\begin{center}
\includegraphics[width=\textwidth]{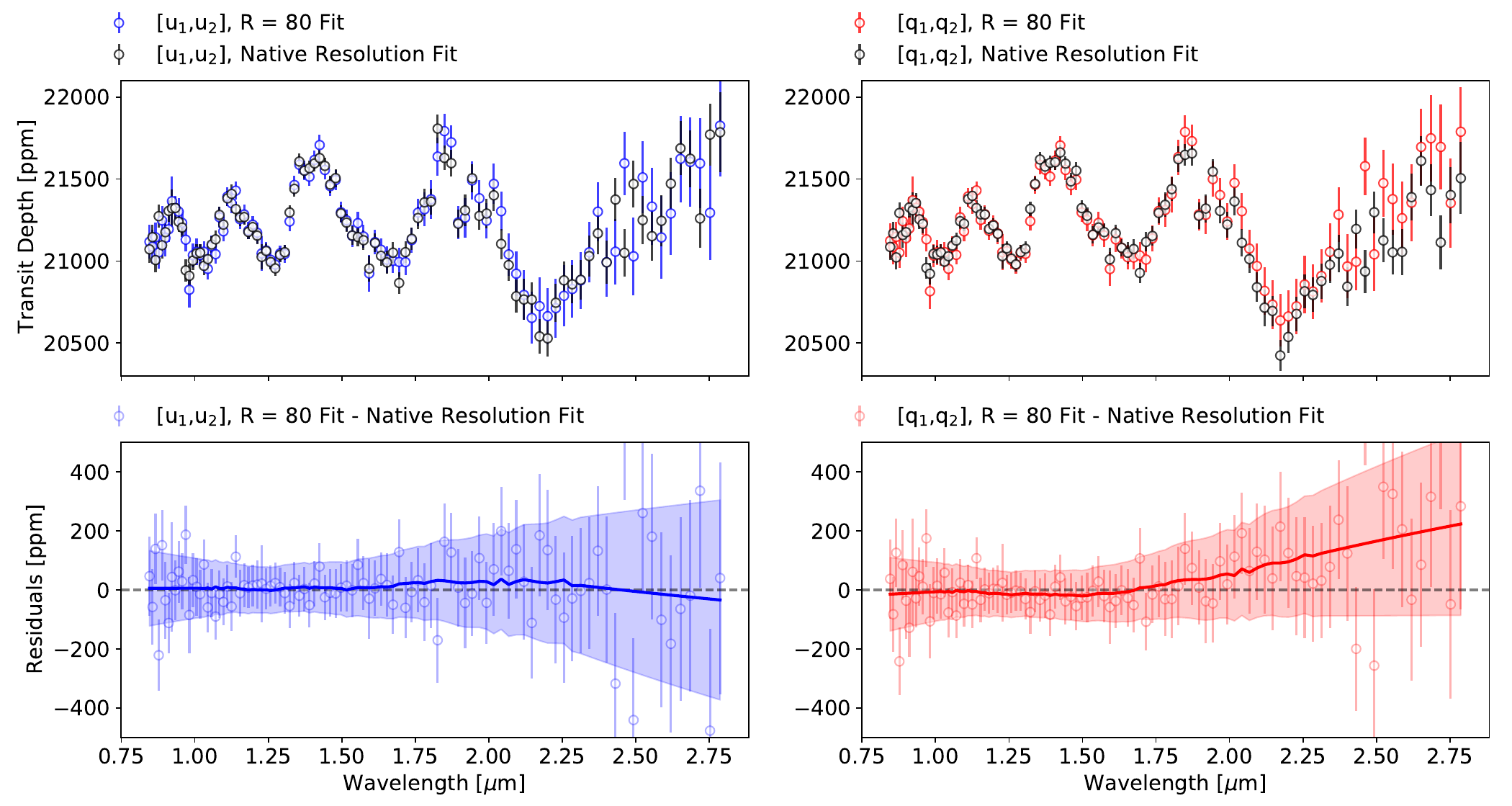}
\end{center}
\vspace{-5mm}\caption{Impact of the choice of limb-darkening parametrization on the measured transmission spectrum. \textbf{Top left:} NIRISS/SOSS spectra of WASP-39\,b, fit at a spectral resolution of R = 80 (blue points) and at native resolution (black points), obtained using the [$u_1$,$u_2$] parametrization with wide uniform priors. The fit performed at native resolution is binned at a resolution of R = 80 after the light curve fitting. \textbf{Bottom left:} Difference between the fit performed at a resolution of R = 80 and the one at native resolution (blue points) for the [$u_1$,$u_2$] parametrization. The error bars for the residuals are computed in quadrature. A smoothed version of the residuals (blue line), along with the 1-$\sigma$ confidence region, are shown using a linear Savitzky-Golay filter with a window size of 31 points. \textbf{Top right:} Same as \textbf{Top left} for the [$q_1$,$q_2$] parametrization. \textbf{Bottom right:} Same as \textbf{Bottom left} for the [$q_1$,$q_2$] parametrization. The residuals between the transit spectra obtained from the light-curve fits at different resolutions clearly show a systematic deviation away from 0 past 2\,$\mu$m in the [$q_1$,$q_2$] case.} 
\label{fig:spectra}
\end{figure*}


\subsection{Simulated data}

We find that, because of the physical assumptions from which the [$q_1$,$q_2$] parametrization is derived, the transit fit is unable to fully and symmetrically explore the parameter space around solutions with low limb-darkening (Figure \ref{fig:lightcurves}). 
To produce a symmetric ensemble of models around a transit light curve in the limit of no limb-darkening, the model must allow for unphysical scenarios with limb brightening, where the stellar intensity profile instead increases from the center to the limb. 
Because the [$q_1$,$q_2$] parametrization is unable to produce such models, the fit instead compensates for this by decreasing the planetary radius (Figure \ref{fig:lightcurves}).

This effect is reminiscent of other astronomical measurements where parameter exploration near a hard boundary or the exclusion of non-physical scenarios results in biased parameter inference. For instance, \citet{Lucy1971} found spectroscopic binaries in circular orbits to be biased towards small non-zero eccentricities due to the rigid boundary at $e$ = 0. This problem is also present in radial velocity fits where, in the limit of low signal-to-noise, negative planetary masses must be allowed in the parameter exploration in order not to bias the measurements \citep[e.g.,][]{radvel2018,eastman2019exofastv2publicgeneralizedpublicationquality}.

The magnitude of the bias on the measured transit depth introduced through the [$q_1$,$q_2$] parametrization scales with the photometric scatter (Figure \ref{fig:scatter}), and reaches approximately 85\,ppm for a photometric scatter of 1000\,ppm. While we keep the current investigation of the bias to its simplest by only fitting for $R_\mathrm{p}$/$R_\star$ and the LDCs, we note that increasing the complexity of the model (e.g., fitting for additional parameters such as $b$ and $a$/$R_\star$) can potentially increase the bias as it provides more flexibility to the transit model. We also find that the magnitude of the bias relative to the measured transit depth uncertainty ($\sigma_\mathrm{t,d}$) does not vary with the light-curve photometric scatter, and is close to $\sqrt{2/\pi}~\sigma_\mathrm{t,d}\approx 0.8~ \sigma_\mathrm{t,d}$. This value corresponds to the mean of a half-normal distribution, consistent with the scenario of a measured parameter whose true value is situated directly at the boundary of the prior space. 

In the context of transmission spectroscopy, this means that a spectrum that is fitted at a high resolution and then binned to a lower resolution will show systematically lower transit depths than a spectrum obtained by fitting light curves that are already binned to the same low resolution. This is because the high-resolution spectrum is more sensitive to the bias since the light curves that are fit have larger scatter. This effect will especially arise at infrared wavelengths where the limb darkening becomes less important. Furthermore, the magnitude of the bias on the measured transit depths for a given spectrum will follow the wavelength-dependent scatter of the light curves, which depends on the instrument throughput and stellar spectrum. The consequence of this effect on JWST transmission spectra has already been observed by \citet{May_2023} and \citet{carter_benchmark_2024}, who then recommended to bin the light curves at a lower spectral resolution before proceeding with the light curve fitting (effectively mitigating the effect of the bias by decreasing the scatter in the light curves) and/or to fix the LDCs to avoid this bias. 

While fixing the LDCs to theoretical model predictions leads to spectra that are consistent no matter the resolution at which they are fit, such models can differ significantly from reality and could in turn bias the measured transmission spectrum \citep[e.g.][]{Espinoza2015,Maxted_2022,Patel2022,Rustamkulov_2023,kostogryz_magnetic_2024,sarkar2024exoplanet}. The continuous temporal coverage and high photometric precision of the JWST allows for a precise measurement of the LDCs from the data and the use of theoretical models should preferably be reserved to planets with high impact parameters, in which case the LDCs and planetary radius are strongly correlated \citep[e.g.][]{Roy_2023,radica2024muted}. We propose instead to use the [$u_1$,$u_2$] parametrization with wide uninformative priors such that the resulting transmission spectrum is free of any bias no matter the spectral resolution used for the fit, which we demonstrate for real observations in the following subsection.

\subsection{NIRISS/SOSS observations of WASP-39\,b}\label{sec:w39}

We find that the bias arising from the [$q_1$,$q_2$] limb-darkening parametrization significantly changes the NIRISS/SOSS transmission spectrum of WASP-39\,b redwards of two microns. The spectra obtained by fitting the light curves at native resolution and at a fixed resolution of R = 80 for both the [$u_1$,$u_2$] and [$q_1$,$q_2$] parametrizations are shown in Figure \ref{fig:spectra}. When the light curves are fitted using the [$q_1$,$q_2$] parametrization, the native resolution spectrum is systematically lower past 2\,$\mu$m than the fit performed on the binned light curves, reaching a discrepancy of up to 200\,ppm at the longest wavelengths. This discrepancy between the spectra fit at different resolutions disappears when using the [$u_1$,$u_2$] parametrization. 

In the case of NIRISS/SOSS, it is mainly the longest wavelengths that are affected by this bias since this is where the scatter is highest and also where the limb-darkening effect becomes negligible. We note that this effect is not significant only for NIRISS/SOSS but for all instruments onboard the JWST due to their coverage of infrared wavelengths. Furthermore, because the instruments onboard the JWST all have different spectral resolutions and throughputs, using the [$q_1$,$q_2$] parametrization may also lead or contribute to offsets between transmission spectra of the same planet observed with multiple instruments. However, such offsets have also been observed in datasets where the LDCs have been fixed \citep[e.g.][]{wallack2024jwst}, meaning other causes might also be at play.

\begin{figure}
\begin{center}
\includegraphics[width=0.45\textwidth]{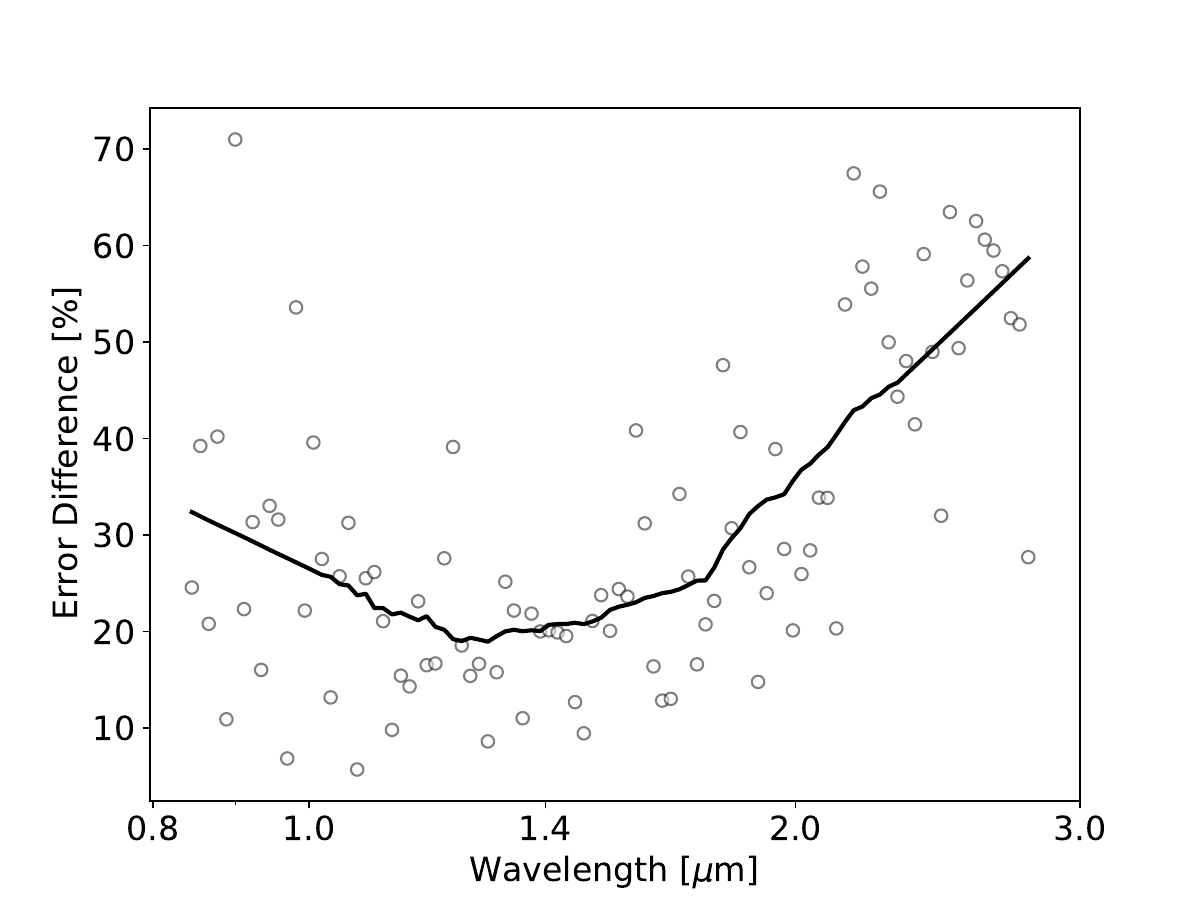}
\end{center}
\vspace{-5mm}\caption{Fitting the light curves at higher spectral resolution produces a more precise spectrum. Relative difference in percent between the errors on the transit depths measured at a resolution of R = 80 and those measured at the native resolution and then fitted afterward. A smoothed version of the difference transit depth precision (black line) is shown using a linear Savitszky-Golay filter with a window size of 31 points. The errors obtained from the lower resolution fit are systematically larger than those obtained at the native resolution, with the difference in precision ranging from 20 to 60\% across the wavelength range.} 
\label{fig:precision}
\end{figure}

We find that the errors on the measured WASP-39\,b NIRISS/SOSS transit depths are 20 to 60\% lower across the spectral order 1 wavelength range for the spectrum where the light curves have been fitted at the native resolution compared to the one that was directly fitted at a resolution of R = 80 (Figure \ref{fig:precision}). We measure this difference in precision in each spectroscopic channel via $$\mathrm{Rel. \ difference}\ = 100\cdot[\sigma_\text{R=80}/\sigma_\text{native}-1]\,\%,$$ where $\sigma_\text{R=80}$ and $\sigma_\text{native}$ are the 1$\sigma$ uncertainties on the spectroscopic transit depths obtained when the light curves are binned before and after the fit, respectively. We find that the gain in precision is most notable towards the blue and red ends of the wavelength range (Figure \ref{fig:precision}), where the flux is lower and thus where there is the most increase in precision to be made.
One possible explanation for this increase in precision is that, as the light curves are binned, the ratio of the systematic-to-Poisson noise $\sigma_\mathrm{sys}/\sigma_\mathrm{Poisson}$ increases, forcing the fitted scatter to compensate for this deviation from pure Poisson noise. However, when we bin the spectrum that was fitted at the native resolution, we assume that the uncertainties decrease following Poisson noise, which could explain the improvement in precision compared to a spectrum that is fitted at a lower resolution. Another possibility is that some of the fit parameters, such as the systematics model parameters and limb-darkening coefficients, introduce uncertainties on the measured transit depth that do not scale following Poisson noise behavior, thus proving advantageous to fit the light curves at a higher spectral resolution. Hence, we conclude that fitting the light curves at the native resolution might be necessary to achieve the best possible precision for transmission spectroscopy.

While we find that a uniform prior $\mathcal{U}[-3,3]$ allows for sufficient exploration of the LDCs parameter space for the simulated data as well as the WASP-39\,b spectrum presented in this work, we note that this range might need to be adjusted depending on the precision of the observations. One should also ensure that the retrieved LDCs are physical and that there is no statistically significant preference for limb-brightening/unphysical scenarios in the light curve fits.

\section{Discussion}\label{sec:discu}

\subsection{Effect on atmospheric inferences}

While the [$q_1$,$q_2$] parametrization bias described above results in measured transit depths that are mostly consistent to one standard deviation with those obtained using the [$u_1$,$u_2$] parametrization (Figures \ref{fig:lightcurves}, \ref{fig:scatter}, \ref{fig:spectra}), its effect on atmospheric inference can be significant. This is because this bias is repeated over many wavelength bins and its magnitude may depend on wavelength for observations where the scatter vary significantly over the bandpass. This bias can thus lead to unphysical trends in the obtained transmission spectra (e.g., slopes, offsets between instruments or detectors) which cannot be reproduced by atmosphere models and must be accounted for with ad-hoc methods such as fitting for detector offsets at the retrieval stage or error inflation. Furthermore, the amplitude of the bias can reach the same order of magnitude as the atmospheric features we are aiming to probe in transmission spectroscopy. Looking at the NIRISS/SOSS spectrum of WASP-39\,b specifically, the bias introduced by the [$q_1$,$q_2$] parametrization reaches 200\,ppm at the longest wavelengths whereas the atmospheric features themselves have an amplitude of approximately 1000\,ppm. Such a systematic downward trend at long wavelengths, affecting practically half of the wavelength range covered by the observations, could impact the inferred atmospheric metallicity, which depends on the relative strength of the water features \citep{Benneke2012}, as well as the need for non-gray clouds and/or hazes to explain the spectrum. This is even more significant for smaller terrestrial planets, at which point the expected amplitude of the transmission spectroscopy features for a secondary atmosphere are on the level of $\sim$100\,ppm and the transit depth precision can reach 25\,ppm (e.g., Benneke et al. submitted)
Performing light-curve fitting of JWST observations with the [$u_1$,$u_2$] parametrization is thus the best way to ensure that the observed transit spectra are free of these limb-darkening-induced biases before proceeding with atmospheric retrievals.

\subsection{Updating the [$q_1$,$q_2$] parametrization}

Because the bias towards smaller planetary radii is introduced by excluding limb brightening from the parameter space and considering only limb-darkening scenarios, one must then ask: is it possible to relax the constraints in the [$q_1$,$q_2$] parametrization such that limb brightening is also allowed? We derive such a parametrization in the Appendix and find that one can sample uniformly from a [$r_1$,$r_2$] unit square to produce the following values of the quadratic LDCs: 

\begin{equation}
u_1 = 2 \left(r_2\sqrt{r_1} + \sqrt{r_1} - 1\right)
\end{equation}

\begin{equation} 
u_2 = \frac{u_1}{|u_1|} \sqrt{r_1} \left(1 - r_2\right) - \frac{u_1}{2},
\end{equation}

\medskip 
\noindent
such that the values of $u_1$ and $u_2$ allowed result only in positive stellar intensity profiles that either monotonically decrease or increase towards the limbs. 
We find however that, other than for the advantage of reducing the size of the prior space and circumventing the bias on the planetary radius, this parametrization results in larger uncertainties on the measured transit depth and might also be trickier to explore by sampling algorithms (see details in the Appendix). It therefore seems that the best option is to simply use the standard [$u_1$,$u_2$] parametrization. 

\section{Summary and Conclusions}\label{sec:conclusions}

We have demonstrated the existence of a bias on the measured planetary radius that is introduced through the use of physically-motivated limb-darkening parametrizations. This bias is most obvious at infrared wavelengths where the limb-darkening effect is low and the assumption of a monotonically decreasing stellar intensity profile hinders a complete exploration of the planetary radius parameter space. Tests on simulated data show that the measured planetary radius is systematically lower when using the [$q_1$,$q_2$] parametrization of \citet{Kipping_2013}. We also show that this bias increases with photometric scatter 
explaining the discrepancy between spectra fit at high versus low spectral resolutions observed in \citet{May_2023} and \citet{carter_benchmark_2024}. This bias can affect large portions of the wavelengths covered by a transmission spectrum, and lead to significant biases in atmospheric characterization currently obtained with the instruments onboard the JWST. 

We find that using the standard [$u_1$,$u_1$] quadratic limb-darkening parametrization with wide uninformative priors prevents this bias and demonstrate this on the JWST NIRISS/SOSS spectrum of WASP-39b. Furthermore, we find that fitting the light curves at the native resolution results in errors on the transit depth that are 20 to 60\% lower across the NIRISS/SOSS wavelength range compared to the spectrum directly fitted at a low resolution, showing that it is potentially necessary to fit at the native resolution of the instruments to maximize the precision of the measured spectrum. 

In the limit of high-precision optical light curves, where the effect of limb darkening is important, and for which such physically-motivated parametrizations were developed \citep[e.g.][]{Burke_2007,Carter2009,Kipping_2013}, the use of the [$q_1$,$q_2$] parametrization (or similar) could still be advisable, as the limb-darkening-free solutions should confidently be excluded from the parameter space. Nevertheless, it is unclear exactly at which wavelengths and what level of photometric scatter the bias on the measured transit depth becomes important. As we move to the analysis of infrared light curves with the JWST and other future space telescopes, the use of such methods should be avoided to ensure that transmission spectra measurements are free of biases. 

\section*{Acknowledgments}
We thank the anonymous reviewer for their insightful comments that improved the quality of this work. We would also like to acknowledge Aarynn Carter, Erin May, Néstor Espinoza, and Luis Welbanks for their role as part of the Early Release Science data synthesis effort in characterizing the discrepancy between JWST exoplanet transmission spectra fitted at low vs high spectral resolution. This work is based in part on observations made with the NASA/ESA/CSA/JWST. The data were obtained from the Mikulski Archive for Space Telescopes (MAST) at the Space Telescope Science Institute, which is operated by the Association of Universities for Research in Astronomy, Inc., under NASA contract NAS 5-03127. The specific observations analyzed can be accessed via
\dataset[DOI: 10.17909/efv0-j030]{https://doi.org/10.17909/efv0-j030}.
B.B. and P.-A.R. acknowledge funding from the Natural Sciences and Engineering Research Council (NSERC) of Canada. L.-P.C. and P.-A.R. acknowledge support from the University of Montréal, and from the Trottier Research institute for Exoplanets (iREx). B.B also acknowledges financial support from the Canadian Space Agency and the Fond de Recherche Québécois-Nature et Technologie (FRQNT; Québec).


\bigskip 
\noindent
\textit{Software:} \texttt{Astropy} \citep{astropy:2013,astropy:2018,astropy:2022}, \texttt{batman} \citep{Kreidberg_2015}, \texttt{emcee} \citep{Foreman_Mackey_2013}, \texttt{Matplotlib} \citep{Hunter:2007}, \texttt{NumPy} \citep{harris2020array}, and \texttt{SciPy} \citep{2020SciPy-NMeth}.

\clearpage

\begin{appendix}

We derive a new parametrization for the limb-darkening, closely following the methodology presented in \citet{Kipping_2013}, where we relax the constraint of monotonically \textit{decreasing} stellar intensity profiles that was used to derive the [$q_1$,$q_2$]. To do this, we impose the two following conditions: 
\medskip 
\begin{itemize}
    \item[\textbf{A.}] The normalized stellar intensity profile cannot reach values below 0 and, per symmetry in the case of limb-brightening, must also not go above 2. 

    \item[\textbf{B.}] The stellar intensity profile must either monotonically decrease or monotonically increase away from the center of the stellar disk.
\end{itemize}

\noindent 
Condition \textbf{A} can be expressed mathematically as the following relation:

\begin{equation}
-1 \leq - u_1(1-\mu) - u_2(1-\mu)^2 \leq 1 .
\end{equation}

\medskip
\noindent 
We can derive the boundary conditions imposed by this relation by evaluating it at its maxima ($\mu$ = 0 and $\mu$ = 1). The $\mu$ = 1 limit returns the trivial condition $-1 \leq 0 \leq 1$. At the $\mu$ = 0 limit however, we have that the standard LDCs must follow the condition $-1 \leq -u_1 - u_2 \leq 1$, from which we can express the boundaries of this parameter space as $u2 = -u_1 - 1$ and $u_2 = -u_1 + 1$. This is consistent with the condition derived in \citet{Kipping_2013} for the case of a positive-everywhere stellar intensity profile.

As for condition \textbf{B}, it can be expressed mathematically as enforcing that the ratio of the gradient of the stellar intensity profile ($\partial I/\partial\mu = u_1 + 2u_2 (1-\mu)$) between any two positions on the stellar disk must be equal to or larger than zero

\begin{equation}
\frac{u_1 + 2 u_2 (1-\mu_1)}{u_1 + 2 u_2 (1-\mu_2)} \geq 0.
\end{equation}

\begin{figure*}[hbt!]
\begin{center}
\includegraphics[width=\textwidth]{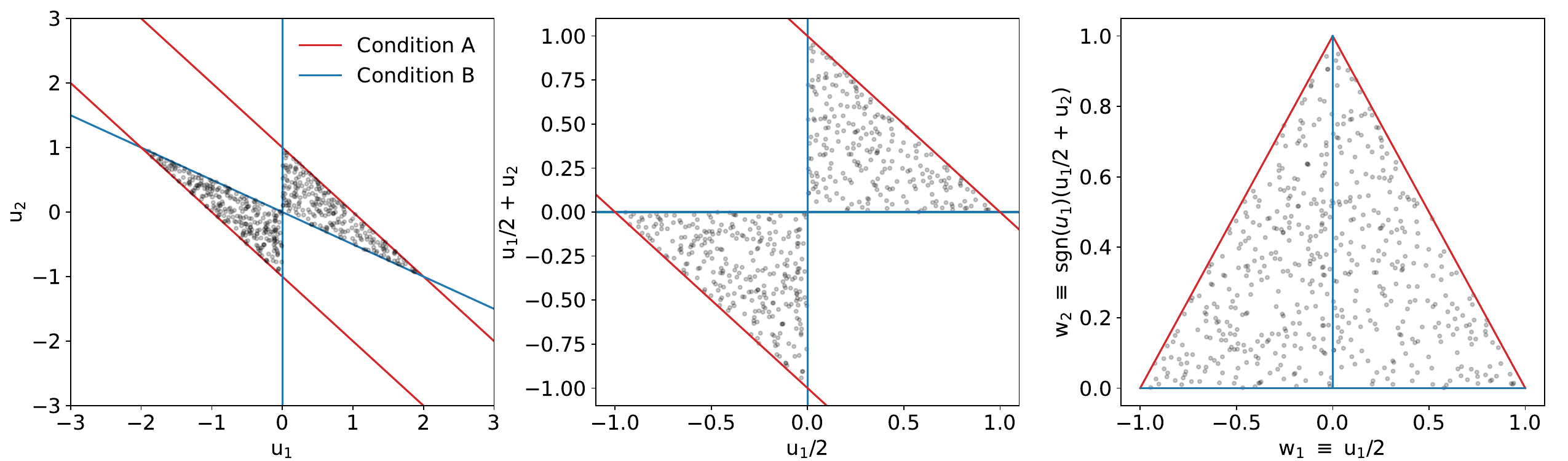}
\end{center}
\vspace{-5mm}\caption{\textbf{Left:} Region of the [$u_1$,$u_2$] parameter space where conditions \textbf{A} and \textbf{B} are respected. \textbf{Center:}. Reparametrization of the limb-darkening parameter space such that the two triangular regions forming the parameter space following conditions \textbf{A} and \textbf{B} are right-angled triangles. \textbf{Right:} Parameter space joining the limb-darkening and limb-brightening scenarios, obtained by rotating the limb-brightening scenarios about the $u_1/2$ axis.} 
\label{fig:conditions}
\end{figure*}

\begin{figure*}[hbt!]
\begin{center}
\includegraphics[width=\textwidth]{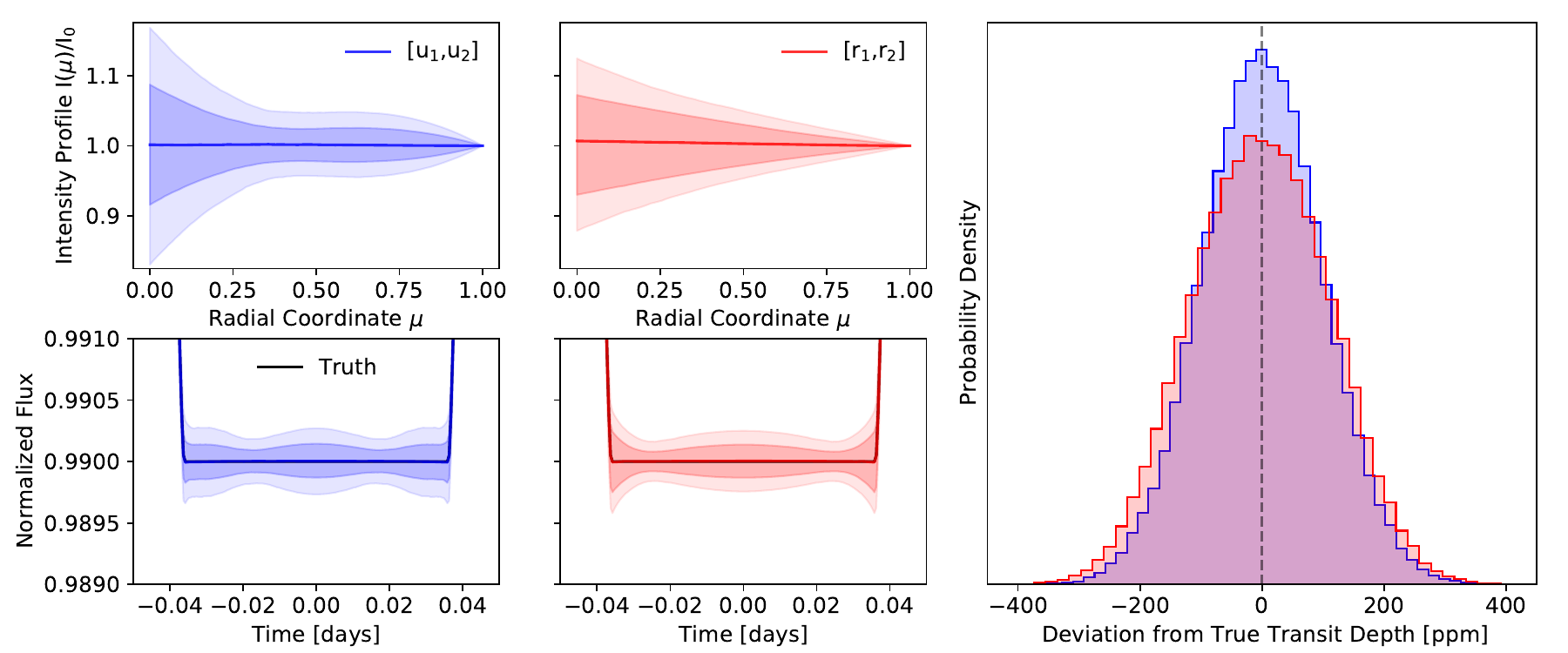}
\end{center}
\vspace{-5mm}\caption{Impact of the choice of limb-darkening on the light curve fit and measured transit depth for the case of no limb darkening and photometric scatter of $\sigma$ = 1000\,ppm. \textbf{Top left:} Median stellar intensity profile $I_\mathrm{s}(\mu)/I_0$ (blue line) from the transit fit to the simulated observations using the [$u_1$,$u_2$] limb-darkening parametrization. The 1$\sigma$ and 2$\sigma$ confidence intervals are shown by the shaded regions. \textbf{Bottom left:} Median transit model (blue) obtained from the transit fit with the [$u_1$,$u_2$] limb-darkening parametrization. The 1$\sigma$ and 2$\sigma$ confidence intervals are shown by the shaded regions. \textbf{Top middle:} Same as \textbf{Top left} considering the [$r_1$,$r_2$] parametrization. \textbf{Bottom middle:} Same as \textbf{Bottom left} considering the [$r_1$,$r_2$] parametrization. \textbf{Right:} Probability density distributions of the measured transit depth minus the true transit depth for the [$u_1$,$u_2$] (blue) and [$r_1$,$r_2$] (red) parametrizations. The transit depths measured using [$u_1$,$u_2$] and [$r_1$,$r_2$] are centered on the expected value, but the posterior distribution obtained with [$r_1$,$r_2$] is wider.} 
\label{fig:r1r2_param}
\end{figure*}

\medskip
\noindent 
By evaluating this equation at its extrema ($\mu_1$ = 0, $\mu_2$ = 1 and $\mu_1$ = 1, $\mu_2$ = 0), we find that the above condition results in the two following boundaries: $u_1\geq0$, $u_2\geq-u_1/2$ and $u_1< 0$, $u_2<-u_1/2$. The regions of the parameter space that follow conditions \textbf{A} and \textbf{B} are shown in Fig. \ref{fig:conditions}. We recover the \citet{Kipping_2013} condition (upper triangle in left panel of Figure \ref{fig:r1r2_param}) in our new bounded parameter space, with the addition of a second region that is now also allowed. This second region corresponds to cases where the stellar intensity profile increases away from the center of the disk (limb-brightening). We then change our system of coordinates such that the x-axis is $u_1/2$ and the y-axis is $u_1/2 + u_2$, transforming our two regions into right-angled triangles. In order to transform this parameter space in a unit square from which we can sample uniformly the limb-darkening parameters, we must join these two regions together into a single triangle. We thus rotate the region that corresponds to limb-brightening scenarios about the $u_1/2$ axis using the parametrization $w_1 \equiv u_1/2$, $w_2 \equiv \text{sgn}(u_1) (u_1/2 + u_2)$. From this, we follow the same method as in \citet{Kipping_2013} and use the triangular sampling of \citet{Turk1990GeneratingRP} such that $(w_1~w_2)^\mathrm{T} = (1-\sqrt{r_1})\vec{A} + \sqrt{r_1}(1-r_2)\vec{B} + r_2\sqrt{r_1}\vec{C}$, where $\vec{A} = (-1~ 0)^\mathrm{T}$, $\vec{B}=(0~ 1)^\mathrm{T}$, and $\vec{C}=(1~ 0)^\mathrm{T}$ are the three vector positions corresponding to the vertices of the triangle in [$w_1$, $w_2$] space. We can then sample uniformly from the [$r_1$,$r_2$] unit square, the values of which can be converted to the quadratic LDCs via the following relations:

\begin{equation}
u_1 = 2 \left(r_2\sqrt{r_1} + \sqrt{r_1} - 1 \right)  ~ ; ~ u_2 = \frac{u_1}{|u_1|} \sqrt{r_1}(1-r_2) - \frac{u_1}{2}.
\end{equation}

\medskip
Having defined our updated limb-darkening reparametrization, we run the same comparison shown in Figure \ref{fig:lightcurves} but this time using the [$r_1$,$r_2$] parametrization, which is shown in Figure \ref{fig:r1r2_param}. While the [$r_1$,$r_2$] parametrization does not seem to induce a bias on the measured transit depth, we find that it actually results in a larger uncertainty on the measurement compared to simply fitting for [$u_1$,$u_2$]. We also derive a second parametrization by rotating the limb-brightening scenarios about the $u_1/2+u_2$ axis and find that it produces virtually the same results as the [$r_1$,$r_2$] parametrization. This increase in the transit depth uncertainty when using [$r_1$,$r_2$] is most likely due to the fact that this parameter space is more difficult to explore. The transition from the limb-darkening to the limb-brightening scenarios happens at the [$u_1=0,u_2=0$] ‘bottleneck' (Fig. \ref{fig:conditions}), meaning the amount of prior space that corresponds to solutions with low limb-darkening/brightening is significantly reduced when using [$r_1$,$r_2$] compared to the [$u_1$,$u_2$] parameter space. 

\end{appendix}

\clearpage

\bibliography{refs}{}
\bibliographystyle{aasjournal}



\end{document}